\documentstyle[aps,prb,multicol,psfig,array]{revtex}
\tighten
\begin{document}

\title{Degenerate ground state in a mesoscopic YBa$_2$Cu$_3$O$_{7-x}$
grain boundary Josephson junction}

\author{E. Il'ichev,$^1$ M. Grajcar,$^{1,2}$
R. Hlubina,$^2$ R.P.J. IJsselsteijn,$^1$ H.E. Hoenig,$^1$
H.-G.  Meyer,$^1$ A. Golubov,$^3$ M.H.S. Amin,$^{4}$ A.M.
Zagoskin,$^{4,5}$ A.N. Omelyanchouk,$^{4,6}$ 
and M.Yu.  Kupriyanov$^7$}

\address{$^1$Department of Cryoelectronics, Institute for Physical
High Technology, P.O. Box 100239, D-07702 Jena, Germany}

\address{$^2$Department of Solid State Physics, Comenius University,
Mlynsk\'{a} Dolina F2, 842 48 Bratislava, Slovakia}

\address{$^3$Department of Applied Physics,
University of Twente, 7500 AE Enschede, The Netherlands}

\address{$^4$D-Wave Systems Inc., 320-1985 W. Broadway, Vancouver,
B.C., V6J 4Y3, Canada}

\address{$^5$Physics and Astronomy Dept., University of British
Columbia, 6224 Agricultural Rd., Vancouver, B.C., V6T 1Z1, Canada}

\address{$^6$B.I. Verkin Institute for Low Temperature Physics and
Engineering, 47 Lenin Ave., 310 164 Kharkov, Ukraine}

\address{$^7$Institute of Nuclear Physics, Moscow State University,
119 899 Moscow, Russia}

\maketitle

\begin{abstract}
We have measured the current-phase relationship $I(\varphi)$ of
symmetric 45$^\circ$ YBa$_2$Cu$_3$O$_{7-x}$ grain boundary Josephson
junctions. Substantial deviations of the Josephson current from
conventional tunnel-junction behavior have been observed: (i) The
critical current exhibits, as a function of temperature $T$, a local
minimum at a temperature $T^\ast$.  (ii) At $T\approx T^\ast$, the
first harmonic of $I(\varphi)$ changes sign.  (iii) For $T<T^\ast$,
the second harmonic of $I(\varphi)$ is comparable to the first
harmonic, and (iv) the ground state of the junction becomes
degenerate. The results are in good agreement with a microscopic model
of Josephson junctions between $d$-wave superconductors.
\end{abstract}

\pacs{PACS}

\begin{multicols}{2}

%
%
%
%
The most important phenomenological difference between the high-$T_c$
cuprates and conventional superconductors regards the orbital symmetry
of the superconducting order parameter.  In the cuprates the pair
potential changes sign depending on the direction in momentum space
according to \cite{vanHarlingen95,Tsuei00} $\Delta (\vartheta )=\Delta
_0\cos 2(\vartheta -\theta )$, where $\vartheta$ is the angle between
the wave vector and the (laboratory) $x$-axis, while $\theta$ is the
angle between the Cu-Cu bond direction of the superconductor and the
$x$-axis.  This unconventional $d$-wave symmetry was predicted\cite
{Sigrist92} and experimentally confirmed\cite{vanHarlingen95,Tsuei00}
to be directly measurable in the Josephson effect between a high-$T_c$
and a conventional superconductor.  Another consequence of the
$d$-wave symmetry is that mid-gap states (MGS) with energy
$\varepsilon=0$ should form on the free surface of a $d$-wave
superconductor if $\Delta (\vartheta)$ has opposite signs on incident
and reflected electronic trajectories. \cite {Hu94} The MGS density
must be maximal for (110)-like surfaces and this prediction has in
fact been confirmed by STM microscopy on YBCO single crystals\cite
{Wei98} which revealed the MGS contribution to the YBCO tunneling
density of states.  The presence of the MGS is expected to influence
in a spectacular way also the Josephson effect in junctions between
$d$-wave superconductors with different crystallographic orientations.
Yet no clear manifestation of the MGS in the Josephson effect in such
junctions has been observed so far, which is a challenge for the
concept of $d$-wave superconductivity in the cuprates.

Moreover, due to possible applications in quantum
computing,\cite{Ioffe99,Blais00} there is substantial interest in
Josephson junctions and circuits with a doubly degenerate ground
state. Such a state was predicted in an asymmetric 45$^\circ$ junction
($\theta_1=0^\circ$ and $\theta_2=45^\circ$, the angles $\theta_{1,2}$
are defined in Fig.~1), since odd harmonics of the Josephson current
$I(\varphi)=\sum_nI_n\sin n\varphi $ are suppressed by
symmetry.\cite{Kashiwaya00,Huck97} The current-phase relation observed
in Ref.~\onlinecite{Ilichev99} indeed showed a substantial
contribution of the second harmonic $I_2$.  However, there is a finite
supercurrent flowing along the interface in the ground state of
asymmetric 45$^{\circ }$ junctions.\cite{Huck97} Therefore they do not
lead to completely quiet qubits in the sense of
Ref.~\onlinecite{Ioffe99}.

\begin{figure}
\centerline{\psfig{figure=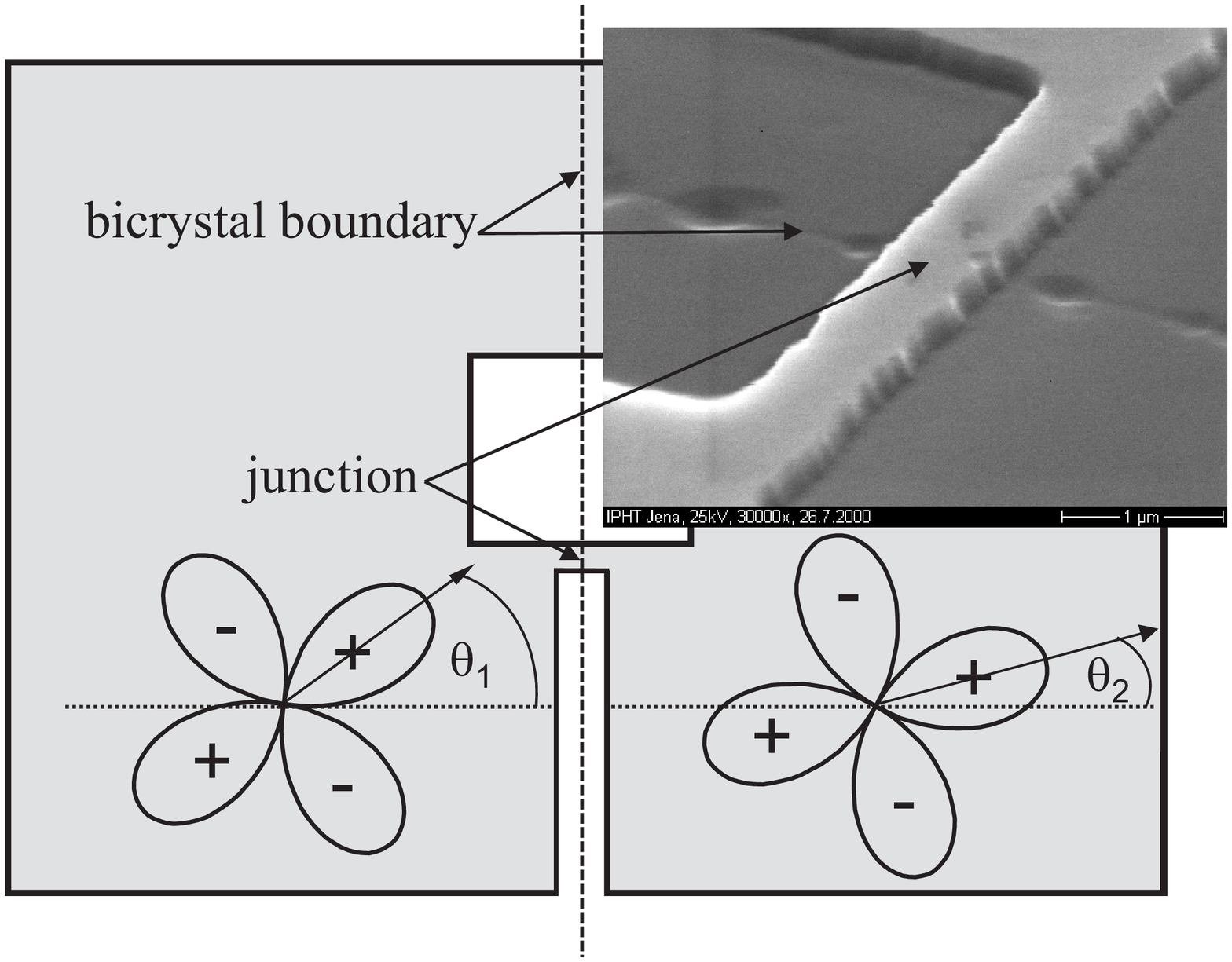,width=7.0cm,angle=0}}
\small{FIG.~1. Schematic picture of the RF SQUID.  The YBCO thin film
occupies the gray area.  The inset shows an electron microscope image
of the narrow grain boundary Josephson junction.}
\label{schemfig}
\end{figure}

Motivated by the search for both, the MGS in high-$T_c$ Josephson
junctions and a quiet qubit, we have studied symmetric 45$^{\circ }$
junctions (i.e. junctions with $\theta _1=-\theta _2=22.5^{\circ
}$). In this paper we report the first direct observation of several
effects exclusive to such junctions: temperature controlled sign
change of the first harmonic of the Josephson current, a nonmonotonic
temperature dependence of the critical current, and the development of
a doubly degenerate ground state of the system.

Let us start with a description of the theoretical predictions for
Josephson junctions between $d$-wave superconductors.  In symmetric
short junctions the Josephson current density is conveniently
described in terms of the Andreev levels in the junction
as\cite{Bagwell98}
\begin{eqnarray}
j(\varphi )=\frac{k_F}{\Phi _0d}\ \sum_n\int_{-\pi /2}^{\pi /2}d\vartheta
\cos \vartheta f\left[ \varepsilon _n(\varphi ,\vartheta )\right] \frac{%
\partial \varepsilon _n(\varphi ,\vartheta )}{\partial \varphi
},\nonumber
\end{eqnarray}
where $\varphi $ is the superconducting phase difference between the
banks, $\Phi _0$ is the magnetic flux quantum, $k_F$ is the Fermi
momentum, $d$ is the average separation of the CuO$_2$ planes,
$f(\varepsilon )$ is the Fermi distribution function, and $\varepsilon
_n(\varphi,\vartheta )$ is the energy of the $n$-th Andreev level for
an electron incident on the junction at an angle $\vartheta $ with
respect to the boundary normal. At a given $\vartheta $ there exist
only two Andreev levels with energies
$\pm\varepsilon(\varphi,\vartheta)$.

The nature of the Andreev levels changes with the impact angle
$\vartheta$: (i) For $22.5^\circ<|\vartheta|<67.5^\circ$, MGS are
formed at $\varphi=0$ whose energy is split by a finite phase
difference $\varphi$ across the contact.  In this range of impact
angles $\varepsilon(\varphi)$ can be qualitatively described by
$\varepsilon_{\rm MGS}(\varphi)=\Delta
(\pi/4)\sin(\varphi/2)\sqrt{{\cal D}(\pi/4)}$, where $0\leq{\cal
D}(\vartheta)\leq 1$ is the angle-dependent barrier
transparency.\cite{Barash00} (ii) For $|\vartheta|<22.5^\circ$ and
$67.5^\circ<|\vartheta|<90^\circ $, no MGS are formed at $\varphi=0$
and the Andreev levels resemble those formed in a Josephson junction
between $s$-wave superconductors. In this range of impact angles
$\varepsilon(\varphi)$ can be qualitatively described by
$\varepsilon_{\rm conv}(\varphi)= \Delta(0)[1-{\cal
D}(0)\sin^2(\varphi/2)]^{1/2}$.
 
When inserted into the equation for $j(\varphi)$, the two sets of
Andreev levels yield contributions of opposite sign to the Josephson
current, $I(\varphi)=I_{\rm MGS}(\varphi)+I_{\rm conv}(\varphi)$.
Close to $T_c$, when $T\gg\Delta_0(T)$, we can approximately write
$I_{\rm conv}\propto{\cal D}(0)(\Delta_0^2/T)\sin \varphi$ and $I_{\rm
MGS}\propto -{\cal D}(\pi/4)(\Delta_0^2/T) \sin\varphi$. For a
sufficiently large ratio ${\cal D}(0)/{\cal D}(\pi/4)$ the sign of the
first harmonic at high temperatures is therefore given by the
conventional contribution.  Lowering the temperature to $\sqrt{{\cal
D}(\pi/4)} \Delta_0(T) \ll T \ll \Delta_0(T)$, $I_{\rm conv}$
saturates to $I_{\rm conv}\propto {\cal D}(0)\Delta_0\sin\varphi$,
whereas $|I_{\rm MGS}|$ continues to grow according to $I_{\rm MGS}
\propto - {\cal D}(\pi/4) (\Delta_0^2/T) \sin \varphi$.  As a result,
near $T^*\sim \Delta_0 {\cal D}(\pi/4)/{\cal D}(0)$ the first harmonic
will change sign and therefore the second harmonic will dominate the
current, leading to a doubly degenerate ground state.

In a symmetric junction $\varepsilon(\varphi,
\vartheta)=\varepsilon(\varphi,-\vartheta)$\cite{Barash00} 
and therefore the total current along the interface is exactly zero
at any $T$ and $\varphi$, contrasting with a finite total current in
an asymmetric junction.  A more detailed analysis shows that in
symmetric junctions time reversal symmetry is broken and small
currents of equal magnitude and opposite sign flow along the interface
in the left and right superconducting banks. Thus, although they are
much closer to being quiet than asymmetric junctions, even the
symmetric junctions are not completely quiet.

%
%
%
%
An experimental confirmation of the above theoretical predictions
requires the use of mesoscopic junctions. In fact, since the grain
boundaries are faceted on a length scale $\sim 0.1$
$\mu$m,\cite{Hilgenkamp96} $I(\varphi)$ of macroscopic junctions
necessarily represents a nontrivial average over junctions with
different misorientations. However, standard transport measurements of
the critical current $I_c=\max_\varphi\{I(\varphi)\}$ are possible
only at temperatures smaller than the energy of the grain boundary
Josephson junction, $\sim\Phi_0I_c/2\pi$. Therefore we have
used the modified Rifkin-Deaver method\cite{Rifkin76,Ilichev98} which
offers a unique possibility to study $I(\varphi)$ at temperatures $T$
much higher than the junction energy. This is achieved by connecting
the banks of the junction to form a superconducting ring (or an RF
SQUID) so that the phase difference across the junction is controlled
by the (large) phase stiffness of the ring.

\begin{figure}
\centerline{\psfig{figure=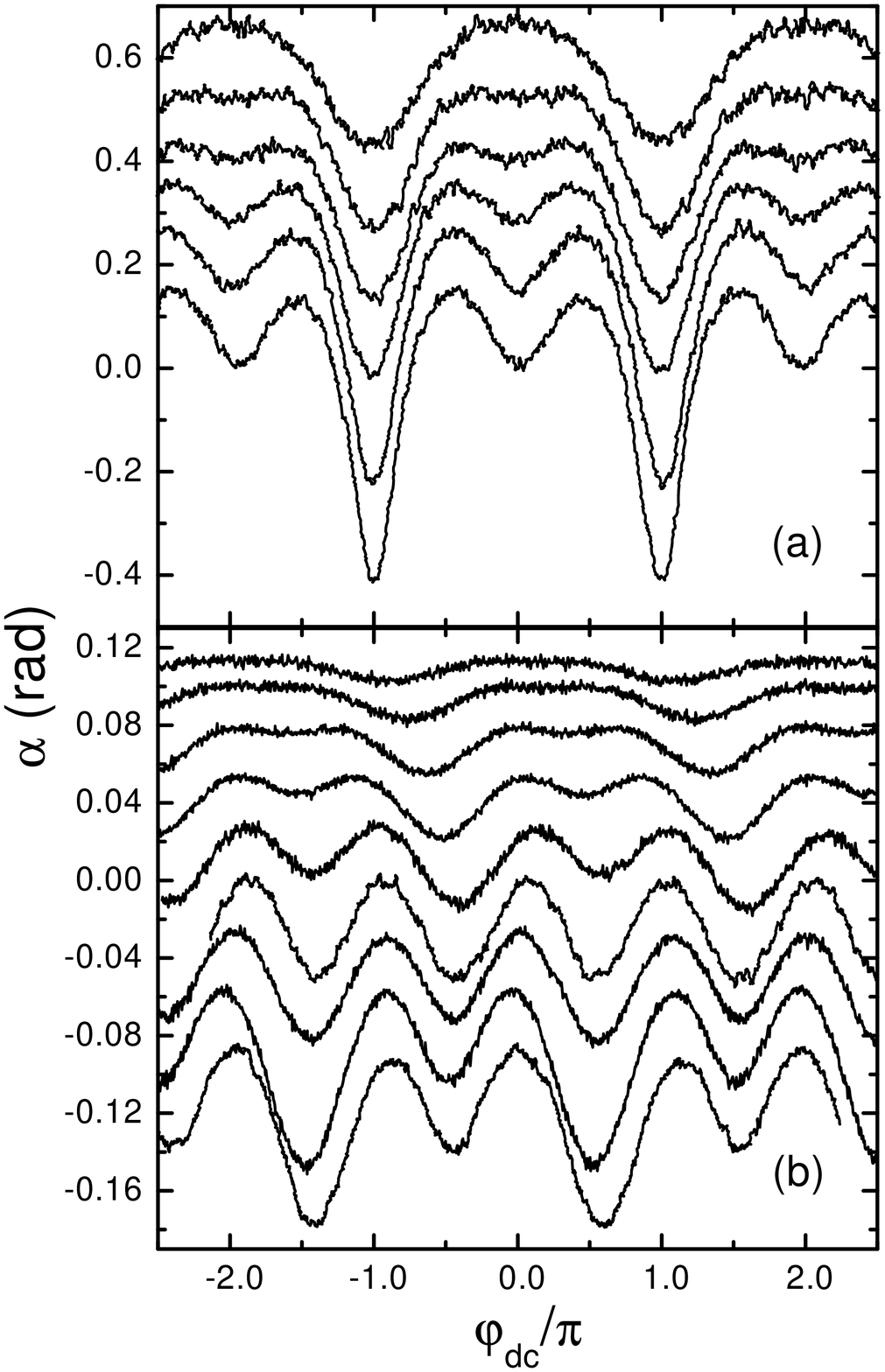,width=6.5cm,angle=0}}
\small{FIG.~2. The phase angle $\alpha$ as a function of
$\varphi_{dc}$ measured at different temperatures for sample No.~1 (a)
and No.~2 (b). From top to bottom, the data correspond to (a) $T$= 30,
20, 15, 10, 4.2, 1.8 K and (b) $T$= 35, 30, 25, 20, 15, 11, 10, 5, 1.6
K.  The data are vertically shifted for clarity.}
\label{alphafig}
\end{figure}

The YBCO thin films of thickness 100 nm were prepared by laser
deposition on 45$^\circ$ symmetric bicrystal substrates. The RF SQUIDs
were patterned in the shape of a square washer 3500 $\mu$m$\times$
3500 $\mu$m with a hole $50$ $\mu$m$\times 50$ $\mu$m by electron beam
lithography (see Fig.~1.). Here we present the data on two samples
with critical current densities $j_c\approx 2.6\times 10^3$ A/cm$^2$
for sample No.~1 and $j_c\approx 400$ A/cm$^2$ for sample No.~2. The
estimated Josephson penetration depth $\lambda_J$ is much smaller than
the width of the wide junction, $w_l=1725$ $\mu$m, and larger than the
width of the narrow junction, $w_s=0.7$ $\mu$m and $w_s=0.5$ $\mu$m
for samples No.~1 and 2, respectively. Thus the behavior of the RF
SQUID is dictated by the narrow junction only. The submicron bridge
was formed at a position between the defects of the substrate which
are visible in Fig.~1.

In the modified Rifkin-Deaver method,\cite{Rifkin76,Ilichev98} the sample is
inductively coupled to a high-quality parallel resonance circuit ($Q=155$
and 165 for samples No.~1 and 2, respectively) driven at its resonant
frequency $\omega_0$. The angular phase shift $\alpha$ between the driving
current and the voltage across the circuit is measured by a RF lock-in
voltmeter as a function of the external magnetic flux $\Phi_{dc}$, which is
conveniently measured in the dimensionless units $\varphi_{dc}=2\pi%
\Phi_{dc}/\Phi_0$. The phase difference across the junction $\varphi$ was
calculated from the $\alpha(\varphi_{dc})$ data using the coupling
coefficient between the RF SQUID and the tank coil $k^2=2.6\times
10^{-3}$ and $3.6\times 10^{-3}$ for samples No.~1 and 2,
respectively. After inverting the $\varphi=\varphi(\varphi_{dc})$
function, $I(\varphi)$ can be obtained from $\beta
f(\varphi)=\varphi_{dc}(\varphi)-\varphi$, where
$f(\varphi)=I(\varphi)/I_c$, $\beta=2\pi L I_c/\Phi_0$, and $L=80$ pH
is the inductance of the RF SQUID. The details of the experimental
method are given elsewhere.\cite{Ilichev98}

\begin{figure}
\centerline{\psfig{figure=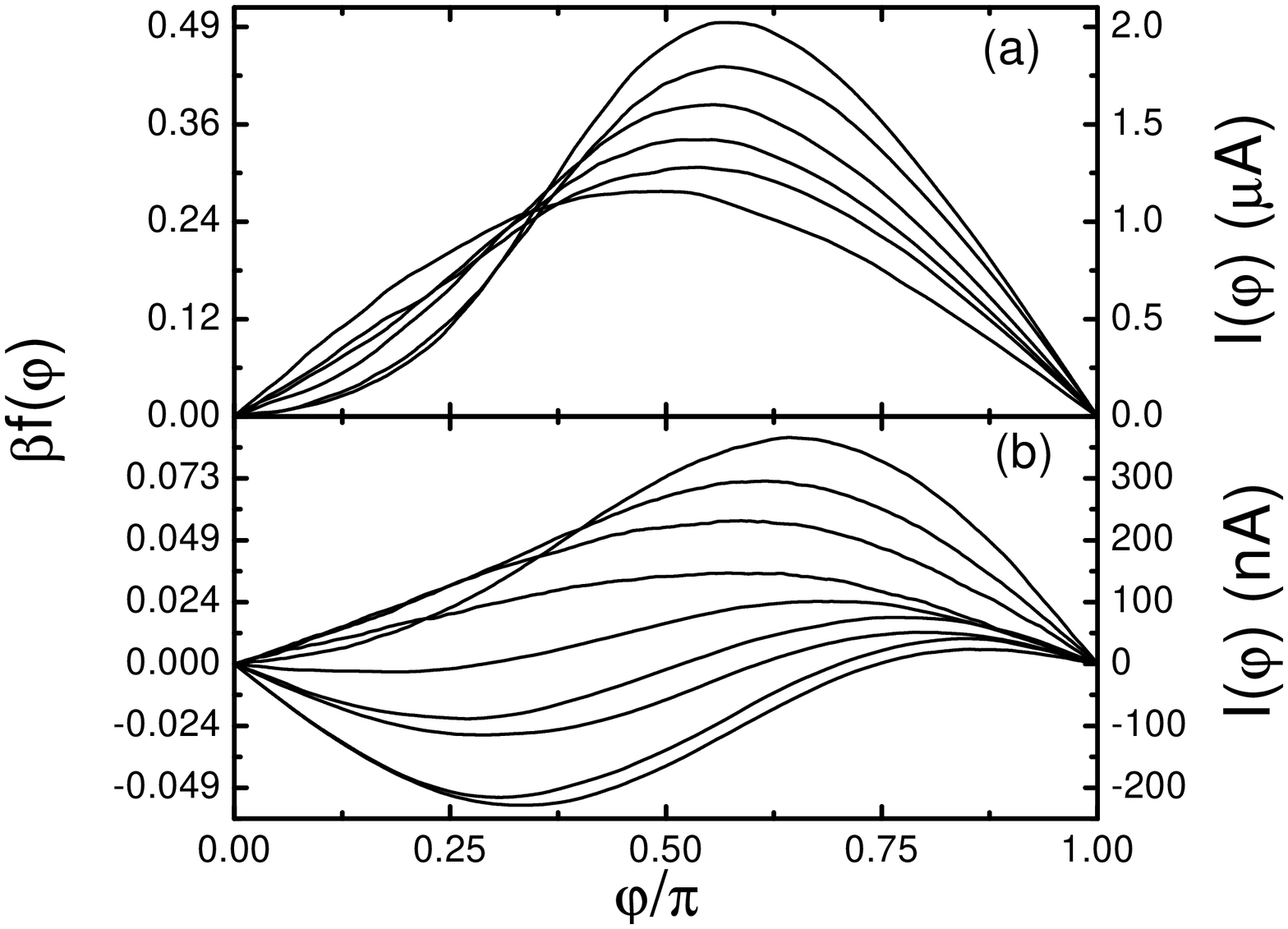,width=7.5cm,angle=0}}
\small{FIG.~3. $I(\varphi)$ for sample
No.~1 (a) and No.~2 (b). From top to bottom at $\varphi/\pi$=0.5, the
data correspond to (a) $T$= 30, 20, 15, 10, 4.2, 1.8 K and (b) $T$=
20, 25, 30, 35, 15, 11, 10, 5, 1.6 K. The data in Fig.~3b
corresponding to $T=20, 25, 30, 35$ were multiplied by a factor 4 for
clarity.}
\label{ipfig}
\end{figure}

The measured $\alpha (\varphi_{dc})$ curves shown in Fig.~2 exhibit
local minima at low $T$ when $\varphi_{dc}$ is a multiple of
$2\pi$. This is qualitatively different from what has been observed
before for $45^\circ$ symmetric grain boundary Josephson junctions on
samples with $w_s>1$ $\mu$m where no such minima were
found.\cite{Ilichev98b} We believe that the difference is caused by
the existence of the bicrystal boundary defects with a typical
distance $\sim 1$ $\mu$m (see Fig.~1), which cannot be avoided for
large junctions.

The Josephson current calculated from the measured
$\alpha(\varphi_{dc})$ data is shown in Fig.~3. Note the anomalous
form of $I(\varphi)$ at low temperatures. The anomalies in sample
No.~2 are much more pronounced than in sample No.~1. We believe this
is a combined effect of smaller junction cross-sections and higher
junction quality, as evidenced by the much smaller values of $I_c$ in
sample No.~2. In Fig.~4 we plot the first two harmonics $I_1$ and
$I_2$ for the sample No.~2. The most striking result is that for
$T^\ast\approx 12$ K, $I_1$ changes sign. In the same temperature
region where $I_1$ starts to exhibit a downturn, the value of $|I_2|$
rises from the negligible high-$T$ values to values comparable to
$|I_1|$ at low $T$, suggesting a common origin of both
phenomena. Furthermore, Fig.~4 shows that close to $T^\ast$, there is
a local minimum of the critical current $I_c$ as a function of $T$,
which is associated with the sign change of $I_1$. These results are
in a qualitative agreement with theoretical predictions for
$I(\varphi)$ of 45$^\circ$ junctions with ideally flat
interfaces.\cite{Tanaka96,Kashiwaya00}

We can reconstruct the free energy $F$ of the junction as a function
of $\varphi$ from $F(\varphi)=(\Phi_0/2\pi)\int_0^\varphi d\phi
I(\phi)$. The result is shown in Fig.~5. Note that for $T\leq 15$ K,
the free energy minimum of the sample No.~2 moves away from
$\varphi=0$, and the $F(\varphi)$ curve exhibits two degenerate minima
at $\varphi=\pm\varphi_0$, as observed previously in
Ref.~\onlinecite{Ilichev99} on asymmetric $45^\circ$ junctions, see
Fig.~5.

%
%
%
%

\begin{figure}
\centerline{\psfig{figure=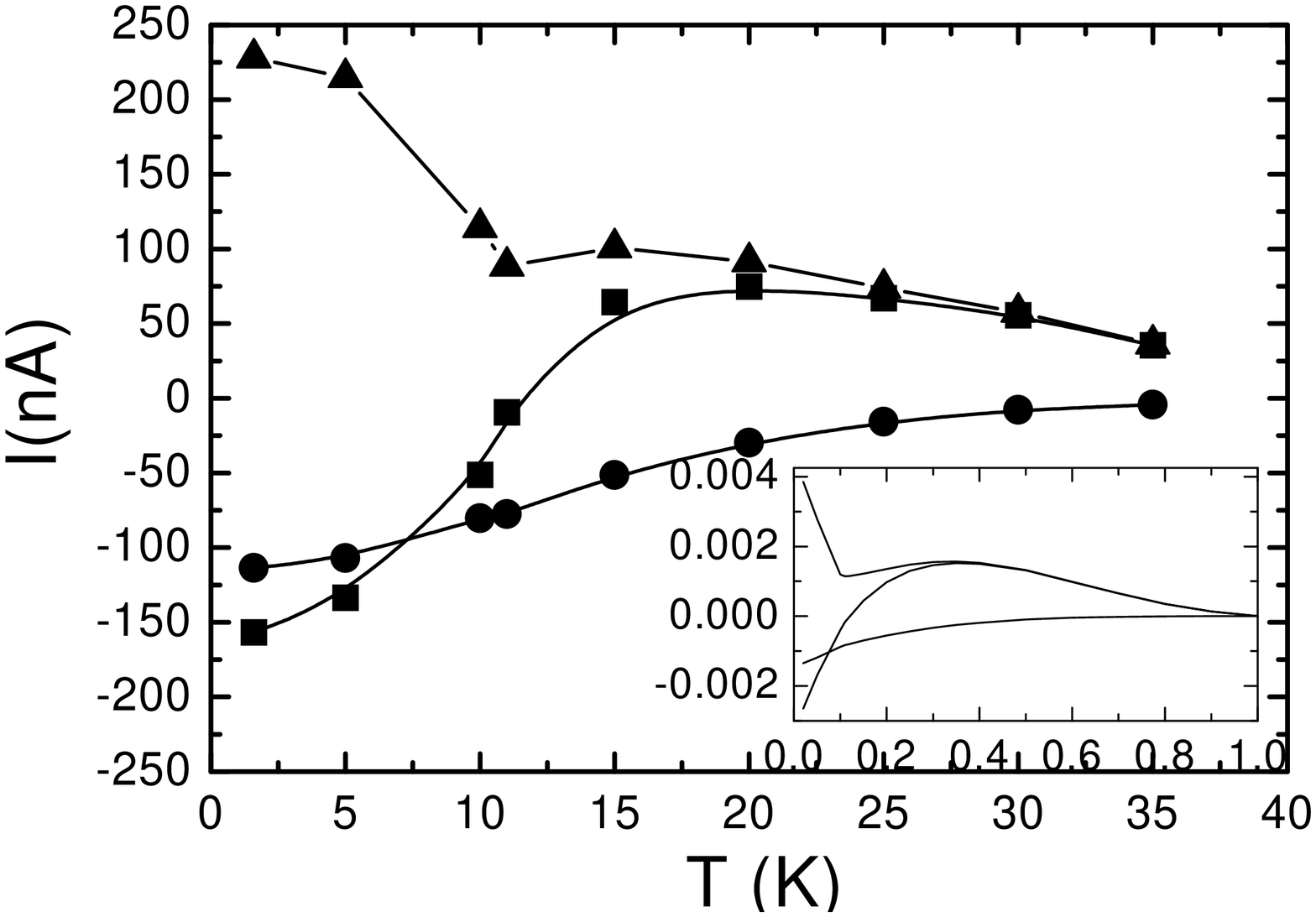,width=7.5cm,angle=0}}
\small{FIG.~4. The critical current $I_c$ (triangles) and the
harmonic components $I_1$ (squares) and $I_2$ (circles) of the
Josephson current as a function of temperature for sample No.~2.  The
figure is obtained by the Fourier analysis of $I(\varphi)$ shown in
Fig.~3b.  Inset: Theoretical prediction for the temperature dependence
of $j_c$, $j_1$, and $j_2$ for a junction with ${\cal D}=0.3$ and
$\rho=0.3$. The current densities are plotted in units of the Landau
critical current density, the temperature is in units of $T_c$.}
\label{i1i2fig}
\end{figure}

In Fig.~4 we compare the experimental data with a theoretical
treatment based on the quasiclassical Eilenberger equations which was
introduced in the $s$-wave case in Ref.~\onlinecite{Omelyanchouk94}
and will be described in detail elsewhere.  Within our approach the
junction is described by two phenomenological parameters, the junction
transparency ${\cal D}$ and the roughness parameter
$0<\rho<\infty$. The temperature dependence of $I_1$, $I_2$, and $I_c$
is fit well by our theory with ${\cal D}=0.3$ and $\rho=0.3$. The
theoretical $j_c$ is reported in Fig.~4 in units of the Landau
critical current density $j_0=k_F\Delta_0/\Phi_0 d$, where $\Delta_0$
is the gap parameter at the interface. The experimental critical
current density $j_c$ for 45$^\circ$ junctions is smaller than
$j_c^{{\rm bulk}}$ by a factor\cite{Hilgenkamp96} $\sim
10^{-5}-10^{-4}$. Remarkably, the absolute value of $j_c$ is also well
described by ${\cal D}=0.3$ and $\rho=0.3$. In fact, in 45$^\circ$
junctions $\Delta_0$ can be estimated from\cite{Hilgenkamp96}
$\Delta_0\sim I_cR_N\sim 10^{-1}-1$ meV, and therefore $j_0/j_c^{\rm
bulk}\sim\Delta_0/\Delta_0^{\rm bulk}\sim 10^{-2}-10^{-1}$.  Together
with the theoretical result $j_c/j_0\sim 10^{-3}$, this explains the
experimental ratio $j_c/j_c^{\rm bulk}$.

\begin{figure}
\centerline{\psfig{figure=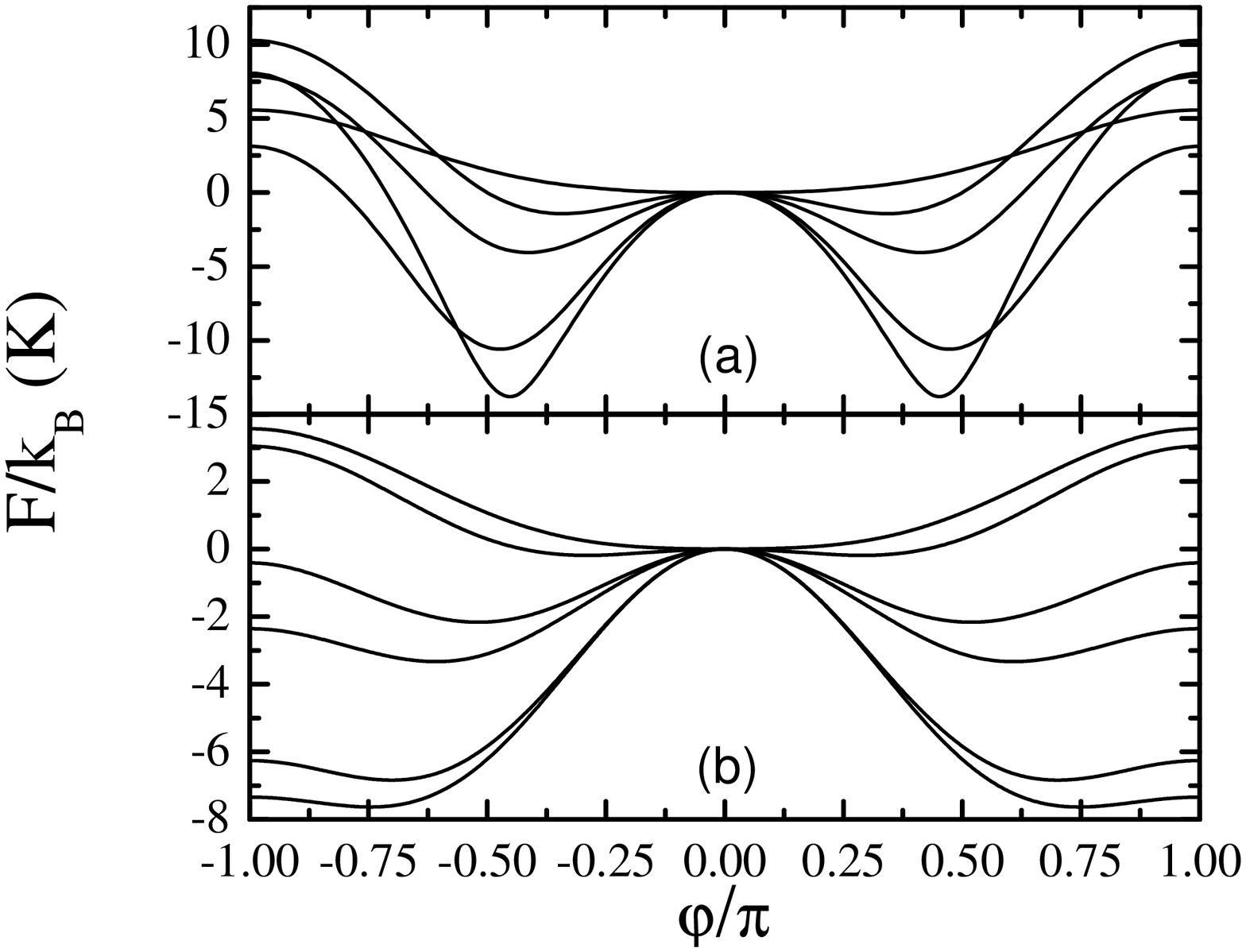,width=7.5cm,angle=0}} \small{FIG.~5.
Free energy $F(\varphi)$ as a function of the phase difference across
the weak link.  The zero of energy has been set so that $F(0)=0$.  (a)
Asymmetric $45^\circ$ grain boundary (Ref.~\onlinecite{Ilichev99}).
Top to bottom curves correspond to $T=$ 30, 20, 15, 10, and 4.2 K,
respectively.  (b) Symmetric $45^\circ$ grain boundary (present work).
Top to bottom curves correspond to $T=$20, 15, 11, 10, 5, and 1.6
K, respectively.}
\label{epfig}
\end{figure}

Before concluding let us point out that a large $I_2$ can be expected
also in a real-space scenario, in which it is postulated that due to
surface roughness and/or twinning of the superconducting banks,
spontaneous magnetic flux is generated along the
interface.\cite{Millis94,Mints98} We are not aware of any prediction
of a sign change of $I_1$ within the real-space scenario. But we can
rule it out also in a different way, on very general grounds: $I_2$
corresponds to a term $-F_2\cos 2\varphi$ in the junction free energy,
where $F_2=(\Phi_0/4\pi)I_2$. In the real-space scenario, $F_2$ is the
free energy gain due to the formation of spontaneous magnetic flux
along the junction. Taking $I_2\approx 120$ nA for the sample No.~2 at
low $T$, we estimate $F_2/k_B\approx 1.3$ K. However, since the
associated pattern of phase difference fluctuations along the junction
is not controlled by the external circuit (unlike the average phase
difference across the junction), thermal fluctuations would smear the
$I_2$ component at $T>F_2/k_B$, in disagreement with experiment.

%
%
%
%
In conclusion, we have found that symmetric $45^\circ$ junctions
exhibit doubly degenerate ground states. Therefore they are of
potential use in superconducting qubit fabrication. The qubits based
on asymmetric 45$^\circ$ junctions are not quiet,\cite{Ioffe99} since
there exists spontaneously generated flux along the
interface\cite{Mannhart96} which should change sign between the two
different ground states of the junction. The qubits based on symmetric
junctions might be close to being completely quiet, since their
spontaneously generated flux is unmeasurably small.\cite{Mannhart96}
The key technological question is how to increase the energy barrier
between the two degenerate minima at $\varphi=\pm\varphi_0$. An
interesting possibility seems to be to increase the barrier
transparency by an appropriate doping of the grain
boundary.\cite{Hammerl00}

Partial support by the DFG and INTAS is gratefully acknowledged.  MG
and RH were supported in part by the Slovak Grant Agency VEGA under Grant 
No.~1/6178/99.

\end{multicols}

\begin{references}
\bibitem{vanHarlingen95}D. J. Van Harlingen, Rev. Mod. Phys. 
{\bf 67}, 515 (1995).

\bibitem{Tsuei00}C. C. Tsuei and J. R. Kirtley, Rev. Mod. Phys. 
{\bf 72}, 969 (2000).

\bibitem{Sigrist92}M. Sigrist and T. M. Rice, 
J. Phys. Soc. Jpn. {\bf 61}, 4283 (1992).

\bibitem{Hu94}C. R. Hu, Phys. Rev. Lett. {\bf 72}, 1526 (1994).

\bibitem{Wei98}J. Y. T. Wei {\it et al.},
Phys. Rev. Lett. {\bf 81}, 2542 (1998).

\bibitem{Ioffe99}L. B. Ioffe {\it et al.}, 
Nature {\bf 398}, 679-681 (1999).

\bibitem{Blais00}A. Blais and A. M. Zagoskin, 
Phys. Rev. A {\bf 61}, 042308-1 (2000).

\bibitem{Kashiwaya00}S. Kashiwaya and Y. Tanaka, 
Rep. Prog. Phys. {\bf 63}, 1641 (2000).

\bibitem{Huck97}A. Huck, A. van Otterlo, and M. Sigrist, 
Phys. Rev. B {\bf 56}, 14 163 (1997).

\bibitem{Ilichev99}E. Il'ichev {\it et al.},
Phys. Rev. B {\bf 60}, 3096 (1999).

\bibitem{Bagwell98}R. A. Riedel and P. F. Bagwell,
Phys. Rev. B {\bf 57}, 6084 (1998).

\bibitem{Barash00}Yu. S. Barash, Phys. Rev. B {\bf
61}, 678 (2000).

\bibitem{Hilgenkamp96}H. Hilgenkamp, J. Mannhart, and B. Mayer, 
Phys. Rev.  B {\bf 53}, 14 586 (1996).

\bibitem{Rifkin76}R. Rifkin and B. S. Deaver, 
Phys. Rev. B {\bf 13}, 3894 (1976).

\bibitem{Ilichev98}E. Il'ichev {\it et al.},
Advances in Solid State Physics {\bf 38}, 507 (1998).

\bibitem{Ilichev98b}E. Il'ichev {\it et al.},
Phys. Rev. Lett. {\bf 81}, 894 (1998).

\bibitem{Tanaka96}Y. Tanaka and S. Kashiwaya, Phys. Rev. B {\bf 53},
R11 957 (1996).

\bibitem{Omelyanchouk94}A. N. Omelyanchouk, R. deBruyn Ouboter, and C. J.
Muller, Low Temp. Physics {\bf 20}, 398 (1994).

\bibitem{Millis94}A. J. Millis, Phys. Rev. B {\bf 49}, 15 408 (1994).

\bibitem{Mints98}R. G. Mints, Phys. Rev. B {\bf 57}, R322 (1998).

\bibitem{Mannhart96}J. Mannhart {\it et al.},
Phys. Rev. Lett. {\bf 77}, 2782 (1996).

\bibitem{Hammerl00}G. Hammerl {\it et al.},
Nature {\bf 407}, 162 (2000).

\end{references}
\end{document}